\documentclass[aip,apl,twocolumn,superscriptaddress,floatfix,amsmath]{revtex4}

\usepackage{graphicx}
\usepackage{dcolumn}
\usepackage{bm}
\usepackage{upgreek}

\begin{document}


\title{Magnonic crystal based forced dominant wavenumber\\ selection in a spin-wave active ring}

\author{A. D. Karenowska}

\email{a.karenowska@physics.ox.ac.uk}
\affiliation{Department of Physics, Clarendon Laboratory, University of Oxford, OX1 3PU, Oxford, United Kingdom}

\author{A. V. Chumak}
\email{chumak@physik.uni-kl.de}
\affiliation{Fachbereich Physik and Forschungszentrum OPTIMAS, Technische
Universit\"at Kaiserslautern, 67663 Kaiserslautern, Germany}

\author{A. A. Serga}
\affiliation{Fachbereich Physik and Forschungszentrum OPTIMAS, Technische
Universit\"at Kaiserslautern, 67663 Kaiserslautern, Germany}

\author{J. F. Gregg}
\affiliation{Department of Condensed Matter Physics, Clarendon Laboratory, University of Oxford, UK}

\author{B. Hillebrands}
\affiliation{Fachbereich Physik and Forschungszentrum OPTIMAS, Technische
Universit\"at Kaiserslautern, 67663 Kaiserslautern, Germany}

\date{\today}

\begin{abstract}
Spontaneous excitation of the dominant mode in a spin-wave active ring --- a self-exciting positive-feedback system incorporating a spin-wave transmission structure --- occurs at a certain threshold value of external gain. In general, the wavenumber of the dominant mode is extremely sensitive to the properties and environment of the spin-wave transmission medium, and is almost impossible to predict. In this letter, we report on a backward volume magnetostatic spin-wave active ring system incorporating a magnonic crystal. When mode enhancement conditions --- readily predicted by a theoretical model --- are satisfied, the ring geometry permits highly robust and consistent forced dominant wavenumber selection.
\end{abstract}


\maketitle

Self-exciting positive-feedback spin-wave systems --- often referred to as spin-wave active rings ---
attract significant attention within the field of magnetic dynamics. The basis of the active ring
is a dispersive spin-wave waveguide with exciting and receiving antennas connected together via a
variable-gain electrical feedback loop. If gain and phase feasibility conditions are met,
noise-initiated signals propagate in the ring, the character of which has a strong nonlinear dependence on the magnetic properties and environment of the waveguide. Such systems not only offer fundamental
insight into nonlinear magnetic dynamics, but provide the possibility of performing practical
investigations into general nonlinear behaviors --- for example, solitonic phenomena, fractal formation
and parametric processes --- not readily observable in other physical domains \cite{kalinikos, mobius, serga, fractals, hagerstrom, ustinov, wu}.

Spontaneous excitation of active ring systems occurs at a threshold value of external gain: the self-generation threshold. At the onset of excitation, the ring signal is
monochromatic, and the spin-wave wavenumber $k_\text{d}$ corresponds to the lowest-loss feasible mode:
the dominant or $k_\text{d}$- mode. If the external gain is increased beyond the self-generation
threshold, the propagation of the dominant mode suppresses the excitation of all others until a second,
higher threshold, beyond which nonlinear splitting processes permit excitations with multiple
$k$-values, and the system becomes multi-moded. The value of $k_\text{d}$ is highly sensitive to the
internal and external magnetic and thermal environment of the spin-wave transmission medium, and --- in
any real system --- is almost impossible to predict. In this letter, we describe an active ring geometry in which, by exploiting spin-wave  reflections from a magnonic crystal, we are able to determinably select the wavenumber $k_\text{d}$ of the dominant mode.

In the presence of an external magnetic field, thin ferri- and ferromagnetic films support several
distinct types of spin-waves \cite{damon}. Owing to their unusual dispersion characteristics
and nonlinear properties, the behavior of backward volume magnetostatic spin-waves (BVMSW) in active
loop systems attracts particular interest \cite{mobius, ustinov}.

\begin{figure}
\includegraphics[width=0.9\columnwidth]{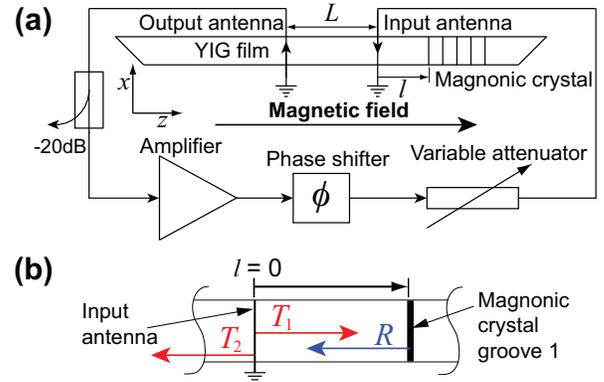}
\caption{\label{Fig1} (Color online). (a) Schematic of experimental active ring
system. Two microstrip
antennas (width 50~$\mu$m) excite and receive spin-waves through a region of YIG film of uniform thickness and length $L=8$~mm. A magnonic crystal (part of the same film) is
located proximal to the input antenna. The film is moveable relative the antennas with precision
$\pm$5~$\mu$m. A bias magnetic field of 160.4~kAm$^{-1}$ (2016~Oe) is applied parallel to the
$z$-axis of the film. (b) Expanded plan ($xz$) view (not to scale) of the region of the YIG
film proximal to the input antenna.}
\end{figure}

In a simple BVMSW active ring --- the conceptual start-point for our study --- two microstrip antennas, separated by a distance $L$, are affixed to a narrow, thin-film spin-wave waveguide magnetized parallel to its long axis. The antennas are connected via an amplifier, a phase shifter (phase shift $\phi$), and a variable attenuator (Fig.~\ref{Fig1}). The combination of the amplifier and the attenuator control the loop gain. Monochromatic modes $k_i$ correspond to feasible, unique, unity values of the Laplace domain transfer function $G(k)$ of the ring: $G(k_i)=1$, at which the loss in the waveguide is exactly compensated by the gain in the
external network (gain feasibility), and the net loop phase shift is zero, or an integer
multiple of $2\pi$ (phase feasibility).  As alluded to above, in a given ring, the dominant mode is ---
in general --- the lowest-loss mode capable of satisfying phase feasibility, but its wavenumber
$k_\text{d}$ is not readily predicted.

\begin{figure}
\includegraphics[width=0.9\columnwidth]{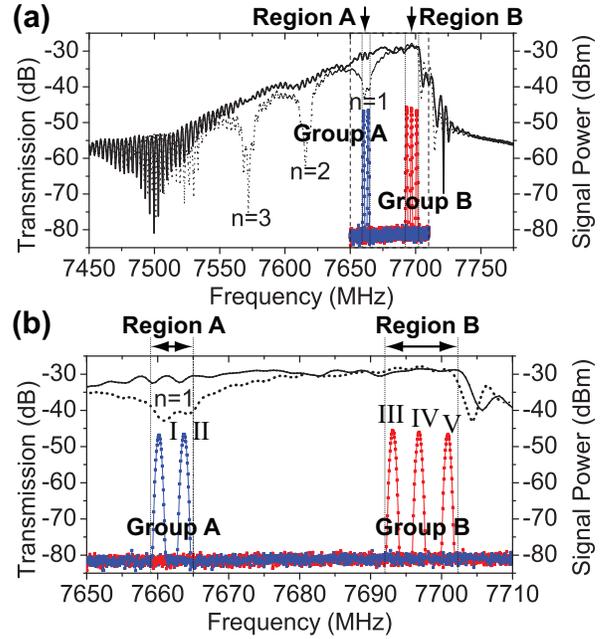}
\caption{\label{Fig2} (Color online). (a) BVMSW transmission characteristics for the
uniform (bold, solid) and magnonic crystal (dotted) regions of the YIG film (left ordinate axis),  $n^\text{th}$ order Bragg
rejection bands of the magnonic crystal are indicated. The FMR frequency is around 7710~MHz. Inset diagram (right ordinate axis) is a composite of five overlaid ring power spectra. Each separate spectrum features a single peak at the frequency corresponding to $k_\text{d}$. Modes positioned within the first rejection band of the magnonic crystal (Region A, Group A) correspond to near-integer values of $l/a$; those in the low-loss region of the spin-wave passband just below the FMR (Region B, Group B) to values outside this range. (b) Magnified view of the section of the upper plot between the dashed lines. Roman numerals relate individual spectral peaks to the modes identified on Fig.~\ref{Fig3}.}
\end{figure}

The active ring system reported in this letter incorporates a magnonic crystal --- a structure with spin-wave transmission characteristics determined by artificially engineered periodic
variations in its geometry --- positioned a variable
distance $l$ from the input antenna (Fig.~\ref{Fig1}(b)) \cite{MC-skyes, Chumak-MC, Chumak-MC2, Chumak-JAP, MC-PRL, MC-APL}. We employ an yttrium iron garnet (YIG) film,
of substantially uniform thickness (5.5~$\mu$m) with a short one-dimensional magnonic crystal region offset from
its center. The crystal comprises 10 grooves of depth 300~nm, width 30~$\mu$m and spacing 270~$\mu$m. The lattice constant is $a=300~\mu$m \cite{Chumak-MC}. The BVMSW transmission properties of this type
of magnonic crystal are well modeled using a treatment similar to that applied to optical and electrical transmission lines \cite{Chumak-MC, Chumak-JAP}. The crystal is helpfully conceptualized as a spin-wave analogue of an optical distributed fibre Bragg grating, with resonant spin-wave reflections from the grooves giving rise to rejection bands. The transmission characteristics of both the uniform region of the film and the magnonic crystal are illustrated in Fig.~\ref{Fig2}. The  first Bragg rejection band of the magnonic crystal (Region A), and the low-loss section of the film's passband between this region and the ferromagnetic resonance (FMR) frequency (Region B) are at the focus of the discussion which follows.

By virtue of the particular symmetry properties of BVMSW, when the ring of Fig.~\ref{Fig1}(a) is active, spin-wave excitations originating from the input antenna propagate both toward the output antenna and the magnonic crystal (complex amplitudes $T_2$ and $T_1$ respectively, Fig.~\ref{Fig1}(b)) \cite{PR-PRB}. The antennas are very narrow in comparison with the spin-wave wavelength. If the wavenumber $k$ of the signal arriving at the magnonic crystal satisfies:

\begin{equation}\label{eq1}
 k= \frac{n\pi}{a},      \quad\quad\quad\quad          \text{where}   \quad n = 1, 2, 3, 4, \ldots
\end{equation}

\noindent (the Bragg condition for the $n^{\text{th}}$ rejection band), the wave
is strongly reflected, and an in-phase signal of complex amplitude $R$ arrives back at the input antenna. The phase accumulation of the reflected signal between
transmission from, and return to, the input antenna is $\phi_{l} = 2 |l| k$. Accordingly, if it is possible to simultaneously satisfy  Eq.~\ref{eq1} and

\begin{equation}\label{eq2}
\phi_{l} = 2 \pi m     \quad\quad\quad\quad \text{where} \quad    m= 0, 1, 2, 3, \ldots.
\end{equation}

\noindent the signal propagating toward the output antenna is enhanced by a factor $H(\rho, l)=(T_2+R){T_2^{-1}}= 1+ \rho e^{-2\beta |l|}$, where $\rho$ is an effective (complex) reflectivity of the magnonic crystal (referenced to $T_2$), and $\beta$ the per-unit-distance transmission loss. This argument suggests that if the magnitude of the enhancement $H(\rho, l)$ is sufficient to increase the effective ring gain in the vicinity of a phase-feasible mode lying in a rejection band of the magnonic crystal (Eq.~\ref{eq1}) above that of all other phase-feasible modes, this mode may be artificially promoted to dominance by arranging that the condition of Eq.~\ref{eq2} is met.

The maximum theoretical augmentation of loop gain attainable through the reflection process is 3~dB. For the spin-wave transmission structure used in the study, only modes associated with the first Bragg order ($n=1$) of the magnonic crystal can be activated through the enhancement mechanism (Region A, Fig.~\ref{Fig2}), since the transmission loss in the unstructured film at $k$-values corresponding to higher orders is at least 10~dB in excess of that associated with the lowest loss region of the spin-wave passband (Region B, Fig.~\ref{Fig2}). Note that the length $L=8$~mm is chosen to balance competing requirements: $L$ must be short so as to minimize the effects of damping and decoherence along the spin-wave transmission path, but long compared with $a$ and --- moreover --- of sufficient length to prevent any reflections from the output antenna back toward the magnonic crystal strongly influencing the ring response. Partial reflection of spin-waves at the output antenna potentially introduces an unmodeled source of phase shift (and effective loss) in the loop, but an analysis of this mechanism is not required to explain the main results of this letter.

For clarity in the discussion which follows, we use the symbol $k^*$ to denote the particular value of spin-wave wavenumber associated with the $n=1$ Bragg reflection in our experimental system: $k=k^*=\pi/a=$104.72~rad/cm (Eq.~\ref{eq1}). A theoretical model predicts an enhancement band associated with $k^*$ centered on 7663~MHz, accessible for integer values of $l/a$ (Eq.~\ref{eq2}).

\begin{figure}
\includegraphics[width=0.9\columnwidth]{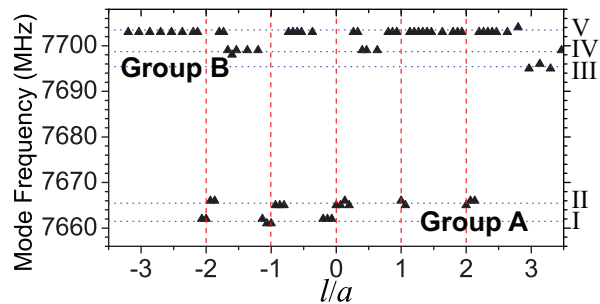}
\caption{\label{Fig3} (Color online). $l/a$ dependence of the $k_\text{d}$-mode frequency
observed at the self-generation threshold.}
\end{figure}

The diagram inset to Fig.~\ref{Fig2}(a) (expanded in Fig.~\ref{Fig2}(b)) is a composite of  overlaid power
spectra obtained via a -20~dB directional coupler within the active ring (Fig.~\ref{Fig1}(a)) for self-generated modes initiated at five distinct $l$ values. At each value of $l$, the external ring gain was increased via the attenuator (Fig.~\ref{Fig1}) from below the self-generation threshold to a value 5~dB above it, and a trace recorded.  Each spectrum features a single peak at the frequency of the corresponding $k_\text{d}$-mode. The sharpness of the spectral peaks shown is limited by the scanning window of the spectrum analyzer, their true width is $\leq25$~kHz. For near-integer values of $l$ ($\pm$1.5\%) the $k_\text{d}$-modes appear within the first rejection band of the magnonic crystal in the vicinity of $k^*$ (Region A, Group A), whilst outside of this range they sit in the lowest-loss region of the pass-band of uniform part of the film (Region B, Group B) at $k$ values determined solely by gain and phase feasibility. Excellent agreement between theory and experiment is indicated. The Roman numerals in Fig.~\ref{Fig2}(b) relate individual spectral peaks to modes identified on Fig.~\ref{Fig3}.

Figure ~\ref{Fig3} shows the $l/a$ dependence of the threshold ring mode for a fixed input $\phi$ at the phase shifter (tuned for phase-feasibility in the vicinity of $k^*$). The vertical dashed lines indicate integer values of $l/a$. The horizontal dotted lines identify discrete quantized modes I to V, falling into lower (Group A) and upper (Group B) frequency bands. The maximum positive integer value of $l/a$ at which forced mode selection is achieved is 2. At $(l/a)\geq3$, loss in the enhancement signal path reduces the loop gain proximal to $k^*$ below that for the lowest-loss modes associated with the uniform region of the film.

Each distinct accessible mode $i$ corresponds to a separate solution of $G(k_i)=1$. Since the phase shift across the spin-wave waveguide is wavenumber dependent, each observed $k$ corresponds to a unique value of phase shift in the external loop. In an ideal system, for fixed $\phi$, we would
expect to observe just two modes: one associated with $k^*$ and integer $l/a$ (the A mode), and one  with general $k$ (the B mode) and non-integer $l/a$ values. The presence of multiple Group B modes is due to the fact that, in the real experimental system, the thickness of the YIG film varies slightly along its length. Accordingly, the dispersion relationship $k(\omega)$ for spin-wave transmission between the antennas depends on the particular 8~mm segment of YIG included in the loop, and thus the value of $l$. This provides a secondary mechanism for the phase accumulation across the waveguide to vary with $l$, independently
of $\phi$.

Forced $k_\text{d}$-mode selection is observed at $l/a = 0$, and extends to negative integer values of $l/a$ (i.e. values of $l$ corresponding to the magnonic crystal entering the region between the antennas). A single mode lying within the $n=1$ Bragg rejection band of the magnonic crystal is associated  with $l/a = +1, +2$. This mode (mode II, Group A) is considered the strict $k^*$-mode. In the vicinity of $l/a = 0, -1, -2$, as well as the strict $k^*$-mode, a second Group A mode (mode I), is observed. The presence of mode I is attributed to a combination of: a) a high effective signal enhancement efficiency associated with $l/a = 0, -1, -2$ loop configurations, b) the effects of non-uniform film thickness discussed above, and c) (specific to $l/a = -1, -2$) bi-directional reflection of spin-waves at the input antenna. Discussion of the interplay between these factors is beyond the scope of this letter, however --- in brief --- it is suggested that in the region of $l/a =0, -1, -2$, there is non-negligible signal enhancement over a range of spin-wave wavenumbers within the $n=1$ Bragg rejection band, including the $k$-value corresponding to mode I, which --- by virtue of the non-uniform thickness effect --- satisfies phase feasibility at certain values of $l/a$.

In summary, a BVMSW active ring geometry incorporating a one-dimensional magnonic crystal has been explored. A wavenumber dependent signal enhancement mechanism reliant on the transmission properties of the magnonic crystal and the geometry of the ring has been shown to enable artificial selection of a dominant feasible ring mode within a very narrow range of a single wavenumber. This wavenumber is associated with the center of the first rejection band of the magnonic crystal and --- as such --- is readily determinable from the crystal's periodicity.

This work serves as a demonstration that combining magnonic crystal and active ring concepts may offer
valuable fundamental insight into spin-wave transmission in structured magnetic films. The active ring arrangement potentially provides a means to investigate scattering phenomena in magnonic crystals, both in linear and nonlinear signal regimes. Additionally, the $k$-value selective mechanism we have established and observed is envisaged to have applications in spin-wave logic structures, microwave oscillators, sensors, and other practical devices.

\medskip

Financial assistance from the DFG SE 1771/1-1 and the DFG GRK 792, and technical support from the Nano+Bio Center, TU Kaiserslautern are gratefully acknowledged. A. K. acknowledges the support of the IET Leslie H. Paddle Scholarship.


\end{document}